\documentclass{IEEEcsmag}

\usepackage[np,newcolumntypes]{numprint}
\npthousandsep{,}
\npproductsign{\times}
\npdecimalsign{.}

\usepackage[square,numbers,sort&compress]{natbib} 
\usepackage[utf8]{inputenc}
\usepackage[english]{babel}

\usepackage{booktabs}
\usepackage{multirow}

\usepackage{url}

\usepackage{array}
\newcolumntype{H}{>{\setbox0=\hbox\bgroup}c<{\egroup}@{}}
\usepackage{subcaption}
\usepackage{soul}
\usepackage[]{mdframed}

\usepackage{tabularx}
\usepackage{nicematrix}

\usepackage{float}
\usepackage{graphicx}

\usepackage{siunitx}
\usepackage{hyperref} 
\usepackage{lmodern}

\usepackage{caption}

\usepackage{standalone}
\usepackage{tikz}
\usepackage{pgf-pie}
\usepackage{pgfplots}
\usepackage{pgfplotstable}
\pgfplotsset{width=6.5cm}
\pgfplotsset{compat=newest}
\usetikzlibrary{shapes, calc}

\let\labelindent\undefined
\usepackage[inline]{enumitem}
\newlist{guidelines}{enumerate}{1}
\setlist[guidelines]{leftmargin=1.68em,label={\sffamily\small G$_{\arabic{guidelinesi}}$}}

\usepackage{soul}

\definecolor{ncsuAquaMild}{HTML}{d9fcec}
\definecolor{ncsuAquaSix}{HTML}{005b5f}
\definecolor{ncsuTeal}{RGB}{0, 176, 183}
\definecolor{ncsuRed}{RGB}{199, 53, 45}
\definecolor{ncsuGreen}{RGB}{124, 185, 43}
\definecolor{ncsuOrange}{RGB}{231, 135, 35}
\definecolor{ncsuPurple}{RGB}{141, 109, 172}

\usepackage{etoolbox}
\usepackage[many]{tcolorbox}

\newcounter{boxno}
\newtcolorbox{mpsFinding}[2][]{%
  colback=ncsuAquaSix,
  colframe=ncsuAquaSix,
  interior style={ncsuAquaMild},
  enhanced,
  title={%
    \refstepcounter{boxno}%
    \ifstrempty{#1}{}{%
      \label{#1}%
    }%
    Finding \arabic{boxno}. #2%
  },
}

\newcounter{textboxno}
\newtcolorbox{mybox}[2][]{%
boxsep=3pt,left=2pt,right=2pt,bottom=5pt,
width=\columnwidth,
boxrule=1pt,
attach boxed title to top center = {yshift=-\tcboxedtitleheight/2},
colbacktitle=white,coltitle=black,
boxed title style={size=normal,colframe=white,boxrule=0pt}, 
interior style={white},
title={\refstepcounter{textboxno}\label{#1}
Example \arabic{textboxno}: {#2}
\def\@currentlabel{\p@textboxno\thetextboxno}},
enhanced,
float,
}

\newcommand{\D}{7} 
 
\newdimen\R 
\newdimen\L 
\newcommand{\A}{360/\D} 

\newcommand{\Menacing}{\textsc{Menacing}}
\newcommand{\Profiling}{\textsc{Profiling}}

\begin{document}

\title{Weapons of Online Harassment:\\ Menacing and Profiling Users via Social Apps}

\author{Sanjana Cheerla}
\affil{NC State University}
\author{Vaibhav Garg}
\affil{Virginia Tech}
\author{Saikath Bhattacharya}
\affil{Illinois State University}
\author{Munindar P. Singh}
\affil{NC State University}

\begin{abstract}
Viewing social apps as sociotechnical systems makes clear they are not mere pieces of technology but mediate human interaction and may unintentionally enable harmful behaviors like online harassment. As more users interact through social apps, instances of harassment increase.

We observed that app reviews often describe harassment.
Accordingly, we built a dataset of \np{3}+ million reviews and \np{1800}+ apps.
We discovered two forms of harassment are prevalent: {\Menacing} and {\Profiling}.

We built a computational model for identifying harassment-related reviews, achieving recalls of \np{90}\% for {\Menacing} and \np{85}\% for {\Profiling}.
We analyzed the reviews further to better understand the terrain of harassment.
Among reviews that mention the abuser’s gender, most identify the abuser as female.
What distinguishes negative from neutral reviews is the greater prevalence of anger, disgust, and fear.

Applying our model, we identified \np{1395} apps enabling harassment and notified developers of the top 48 with the highest user-reported harassment.

\end{abstract}

\maketitle

\textcolor{blue}{This article has been accepted for publication in Computer. This is the author's version which has not been fully edited and content may change prior to final publication. \textbf{Citation information: DOI 10.1109/MC.2025.3587710}}

\section{Introduction} 
\label{sec:introduction}
Unlike traditional social media websites, mobile apps maintain a constant presence in users' lives through phone notifications and easy access. 
We focus on \emph{social apps}, which mediate user interactions, enabling positive interactions and antisocial behaviors such as harassment. 
Social apps span diverse areas like social and professional networking, chatting, dating, photo and video sharing, and anonymous forums.
Importantly, we view an app not as software but as a sociotechnical system that encompasses users and their interactions.

\emph{Online harassment} occurs when one user (abuser) acts through an online platform (here, an app) to interfere in the life of another user (victim) or otherwise intimidate them \citep{un_online_harassment}. 
Most existing works on online harassment don't consider mobile apps and instead focus on social media websites. 

Online harassment is widespread. 
According to the Pew Research Center, \np{41}\% of US adults have experienced it---\np{25}\% via messaging apps, and \np{10}\% via dating apps \citep{vogels2021state}. 
For example, an abuser can use a mobile app to monitor or control the victim's phone, e.g., to track their whereabouts \citep{almansoori2022global}. 

Interestingly, app reviews often describe such harassment and offer insights into victims' concerns. Pater et al. \citep{v1_policies_social_media_platforms} emphasize user feedback in developing harassment policies and norms and app reviews represent actionable such feedback. For example, Malgaonkar et al. \citep{app_review_user_concern} prioritize bug-fixing based on app reviews and Garg et al. \citep{Garg_CACM_audits} use app reviews to facilitate mobile app audits.

Our research objective is to analyze user-reported harassment on social app reviews to foster safer online spaces. Therefore, we address the following research questions.
\begin{itemize}
    \item[] \textbf{RQ$_1$} What types of harassment do users report in social app reviews?
    \item[] \textbf{RQ$_2$} What emotions do victims express regarding harassment in app reviews?
\end{itemize}

Previous analyses focus on app metadata, which is produced by developers and doesn't reveal harassment. In contrast, we mine social app reviews to get the users' perspective. We further contribute a high-recall classifier to detect two types of harassment in app reviews.

By considering \np{1000} reviews, we identified two prevalent types of harassment---{\Menacing} and {\Profiling}. 
{\Menacing} involves sending unwanted or aggressive messages to the victim, where the communication is the attack. {\Profiling} involves collecting or misusing the victim’s personal information to deceive, exploit, or threaten them.
These types roughly follow a United Nations report \citep{un_online_harassment}, though we incorporate ``outing and trickery'' and ``denigration'' into {\Profiling} to account for behaviors such as doxxing that are intended to harm a user's reputation..

\begin{LaTeXdescription}

\item[{\Profiling}] includes the abuser gathering or mishandling the information about the victim:
\begin{enumerate}
    
    \item \emph{Stalking}: attempting to discover personal details through invasion of privacy.

    \item \emph{Predation}: using fake profiles to collect the data and use that data to groom or scam people (primarily on other platforms):
        \begin{enumerate}
        \item Grooming the victim for an ulterior motive.
        \item Scamming for money or similar assets.
        \end{enumerate}
    
    \item \emph{Blackmailing}: using the gathered information from the app to extorting a victim (primarily on other platforms)
    
    \item \emph{Doxxing}: publishing personal information without consent.
\end{enumerate}

\item[{\Menacing}] includes abusers sending unsolicited messages to the victim. Specifically:
    \begin{enumerate}
        \item \emph{Pestering}: nonconsensual intimate images of themself, others, or the victim. 
        \begin{enumerate}
            \item Offering sexual favors for money.
            \item Sexual messages sent without consent. 
        \end{enumerate}
        \item \emph{Child Exploitation}: Attempting pedophilia and child abuse.
        \item \emph{Bullying and Trolling}: insulting or disparaging the victim based on their appearance, gender, or race.
        
    \end{enumerate}
    
\end{LaTeXdescription}

Our work includes all instances of threats of physical or sexual violence and unsolicited sexual content targeting both adults and minors. We exclude cases where sexual content is consensually shared between adults.

\section{Related Work}
\label{sec:related_work}
This section focuses on previous works addressing (i) technology-facilitated online harassment, (ii) the identification of disturbing language in app reviews as a form of harassment, and (iii) the violation of regulations by mobile apps.

Almansoori et al. \citep{almansoori2022global} discuss technology-facilitated intimate partner violence, where abusers use digital tools, including mobile apps, to monitor, control, and intimidate their victims.  Similarly, Garg et al. \citep{Garg_CACM_audits} identify stalking or spying apps (a subcategory of {\Profiling}) from app reviews. However, these studies focus only on a narrow kind of harassment, whereas we consider a general form of harassment prevalent on social apps.

Lin et al. \citep{Lin2021} develop a database of removed iOS apps and categorize reasons for removal, such as ranking, description, and in-app purchase fraud. However, their dataset does not consider user misbehavior and reviews.  Wang \citep{wang2018} categorizes apps as malicious, privacy-risk, fake, spamming, and privacy-violating but overlooks their reasons for removal.

Some researchers \citep{IPVspyware2018, kamal_hatebase} address spyware or hate speech but not online harassment. Roundy et al. \citep{Creepware2020} rely on app metadata such as app descriptions and installation information instead of user reports. 
Freed et al. \citep{Freed2019} analyze online forums where abusers discuss IPS strategies and techniques without considering other platforms where victims or abusers discuss their experiences.

Some researchers investigate {\Profiling} through mobile apps. 
Kumar et al. \citep{kumar2021designing} analyze abusive online behavior on social media. Farnden et al. \citep{farnden2015privacy} study privacy risks in mobile dating apps and note that targeted robbery and sexual assault cases are tied to the use of these services. They also identify malicious artifacts containing sensitive and personally identifiable information.  Traditional mobile app audits consist of source code reviews \citep{EffectiveAudits2019}, neglecting misuses from user actions which we cover in our study. 

\begin{mybox}[box:harassment-relevant]{Online harassment cases.}
\textbf{{\Menacing}: Sending intimate images and bullying}\\
``I've been logged out of my accounts countless of times because I'm simply not interested in the people that text me so they report me when I leave them on seen. So many men are sending me nudes or bullying me on this app. This app is terrible don't use it\ldots''
\\

\textbf{{\Menacing} \& {\Profiling}: Asking for intimate images and personal information} \\
``This app is filled with disgusting creeps that ask for your Snapchat twitter Facebook and insta they will ask for a picture of you deny it they will ask for your name deny it and they will ask for your nudes please deny it\ldots''
\\

\textbf{{\Menacing}: Sending sexual messages}\\
``This is the absolute worst app I have ever been on. The rape culture is unreal. A guy just commented on a post of mine saying he wanted to beat and rape me\ldots''
\\

\textbf{{\Profiling}: Stalking activity}\\
``\ldots I had a person cyberstalk me for weeks and \ldots I feared for my life and you get off on a technicality. I'm going to the Police tomo3rrow to report my cyber stalker. I wish SKOUT actually cared about it's user as it claims it does.''
\\

\textbf{None:}\\ ``It doesn’t work on Apple Watch only on phone so not what I was looking for''
\end{mybox}

\section{Dataset and Model Training}
\label{sec:dataset}
Our dataset collects social app reviews from Apple App Store and Google Play Store. Labeled reviews from the Apple set were used to train the computational model, which we used to analyze both sets. 

We took  193 ``popular'' social apps from Apple\footnote{https://apps.apple.com/us/genre/ios-social-networking/id6005} and identified similar apps. 
We retrieved 692 apps and over 1.6 million negative and neutral reviews. 

We computationally selected 92 seed apps for Google\footnote{Google Play Store: https://play.google.com/store/apps} and added similar apps to obtain \np{1182} apps and 1.8 million negative and neutral reviews.

We developed a multilabel classification model with the following classes: {\Menacing}, {\Profiling}, both {\Menacing} and {\Profiling}, and Neither.

For model training (detailed in the supplement), we selected \np{3050} reviews from Apple matching our keywords; two annotators labeled them. Then, following active learning, we iteratively identified reviews from Apple that didn't match our keywords and labeled those our model had low confidence on, producing a total of \np{7050} manually labeled negative and neutral reviews. 

Next, we trained several machine learning models using the keyword-matching reviews and selected the best-performing model (XLNet). Then, we used the remaining reviews for three rounds of active learning to minimize keyword bias. We trained our computation model using all labeled reviews (\np{7050} reviews) with five epochs of stratified five-fold cross-validation. Finally, we applied the trained XLNet model to both Google and Apple reviews to obtain our results (Section~\ref{sec:results_and_analysis}).

Our computational model achieves high recall: 90\% for {\Menacing} and 85\% for {\Profiling}. We prioritize recall to ensure harassment cases are not missed, even though it leads to increased false positives and lower precision. We make this trade-off since missing true harassment cases is worse than sometimes incorrectly classifying benign reviews.

Example~\ref{box:harassment-relevant} presents snippets of reviews. The first review indicates intimate images and bullying ({\Menacing}), but there are no indications of {\Profiling}. The second review indicates the abusers trying to get intimate images and personal information; thereby, it is included in both categories. The third review indicates {\Menacing} because of sexual messages and no indications of {\Profiling}. The fourth review includes stalking ({\Profiling}) and no indications of {\Menacing}. The fifth review lacks indications of both categories.

We assumed reviews were truthful and that ambiguous actions were user-initiated. We excluded cases where adults consensually shared sexual content but included any instance of an underage user doing so. Furthermore, we flagged activities that could lead to such cases, such as requests for explicit images.

The annotators discussed labeling mismatches to reach agreement. Two annotators labeled a total of \np{7050} reviews with a Cohen's Kappa score of \np{0.81} for {\Menacing} and \np{0.87} for {\Profiling}, which is above the excellent threshold for Cohen's kappa \citep{cohens_kappa_significance}. High inter-annotator agreement ensures reliable labels and improves model training quality by reducing noise and confirming well-defined harassment categories.

\section{Results and Analysis}
\label{sec:results_and_analysis}
We applied our model to the remaining \np{3.4} million reviews in our dataset and identified \np{115581} reviews and \np{1395} apps belonging to {\Menacing}, {\Profiling}, or both.
Figure~\ref{fig:stackedbarchart_overall} shows some distributions we observed.

Finding~\ref{finding:RQ1} is the main outcome for RQ$_1$ on the prevalence of harassment.

\begin{mpsFinding}[finding:RQ1]{Harassment in Social App Reviews}
{\Profiling} is more prevalent than {\Menacing} across both app stores, with Google showing 69.8\% {\Menacing} and 27.2\% {\Profiling} reviews, while Apple shows 44\% {\Menacing} and 30\% {\Profiling} reviews.
\end{mpsFinding}

Both Apple and Google show a higher {\Profiling} prevalence, suggesting that it either occurs more often or is reported more frequently than {\Menacing}.

\begin{flushleft}
\begin{figure}
    \centering
    \begin{tikzpicture}
        \begin{axis}[
            axis line style = {draw=none},
            ybar stacked,
            bar width=0.6cm,
            enlargelimits=0.1,
            legend style={at={(0.5,-0.45)},
            anchor=north,legend columns=-1},
            ylabel={\% of Reviews},
            symbolic x coords={Google, Apple, Male Abuser, Female Abuser, Negative, Neutral},
            xtick=data,
            xticklabel style={yshift=5pt},
            xtick style={draw=none},
            x tick label style={rotate=45, anchor=east},
            ymin=0,
            ymax=100,
            ytick={0,20,...,100},
            ]

            \addplot+[ybar, fill=ncsuGreen, draw opacity=0] plot coordinates {(Google, 69.8) (Apple, 44) (Male Abuser, 51) (Female Abuser, 40) (Negative, 65.7) (Neutral, 51.8)};
            
            \addplot+[ybar, fill=ncsuRed, draw opacity=0] plot coordinates {(Google, 27.2) (Apple, 30) (Male Abuser, 49) (Female Abuser, 60) (Negative, 26.5) (Neutral, 41.8)};
            
            \addplot+[ybar, fill=ncsuTeal, draw opacity=0] plot coordinates {(Google, 3) (Apple, 26) (Male Abuser, 0) (Female Abuser, 0) (Negative, 7.8) (Neutral, 6.4)};
            \legend{\Profiling, {\Menacing}, Both}
        \end{axis}
    \end{tikzpicture}
    \caption{Distribution of reviews that are {\Profiling}, {\Menacing}, or both across various categories.}
    \label{fig:stackedbarchart_overall}
\end{figure}
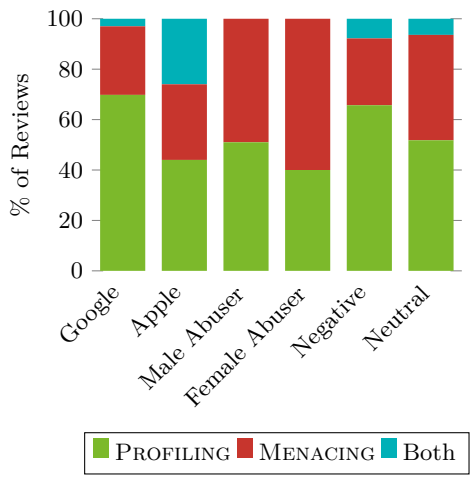
\end{flushleft}

\subsection{Abusers' Genders} 

Figure~\ref{fig:stackedbarchart_overall} male abuser and female abuser display how male and female abusers are reported. Male abusers are equally reported in {\Menacing} and {\Profiling} reviews and female abusers are more often reported in {\Profiling} than {\Menacing} reviews.

\begin{mpsFinding}[finding:gender]{Abuse Strategies by Gender}
Female abusers more often exploit users by gaining monetary assets, and male abusers attempt to obtain personal information or intimate images from users. 
\end{mpsFinding}

We employed an unsupervised method to extract the abuser's gender from reviews explicitly mentioning gender (additional information in the supplement). Out of our \np{115581} reviews, it extracted gender information from \np{18459} reviews (approximately 17\%). We focus on the gender revealed by the user.

For {\Menacing} (\np{8078} reviews), we found that 58\% of the abusers are female and 42\% are male. For {\Profiling} (\np{10381} reviews), 68\% of abusers are female and 32\% are male. 

There are 10\% more female abusers in {\Profiling} than {\Menacing}. Example~\ref{box:male_vs_female_abusers} provides reviews of female and male abusers, highlighting the types of harassment and actions the abusers performed.

\begin{mybox}[box:male_vs_female_abusers]{Female vs. Male abusers.} 
\textbf{Male Abusers} \\
\Menacing: ``I got a PM from a guy that was probably in his $50$s (I’m a minor) trying to flirt with me\ldots'' \\ \\
{\Menacing}: ``I had more than one guy ask me for nudes try to expose my body on the internet\ldots I’ve almost ended my life because of it''\\ \\
{\Profiling}: ``I received a text from a man I never even talked from this app How did he get my number''\\

\textbf{Female Abusers} \\ 
\Menacing: ``I had to report her for sexual content for spam.''\\ \\
{\Profiling}: ``Unfortunately almost every girl I got a message from was trying to\ldots get me to buy things for her'' \\ \\
{\Profiling}: ``There are a lot of girls trying to get OF [OnlyFans] subscribers IG [Instagram] followers etc\ldots''
\end{mybox}

Our qualitative analysis of these posts reveals that female abusers engage in victim profiling more frequently by attempting to obtain personal information and monetary assets.
We illustrate such cases in Example~\ref{box:popular_app_harassment_examples}. In {\Menacing} and {\Profiling}, female abusers send inappropriate images or messages to the victims with links to external, often pornographic, scam sites. Conversely, male accounts attempt to acquire inappropriate images in {\Menacing} and victim's personal information in {\Profiling}.

\subsection{Victims' Emotions}
Analyzing user emotions helps identify their expectations and guides developers in focusing on important concerns.
Negative app reviews that express surprise indicate that users encountered unexpected problems, helping developers address specific problem areas. 

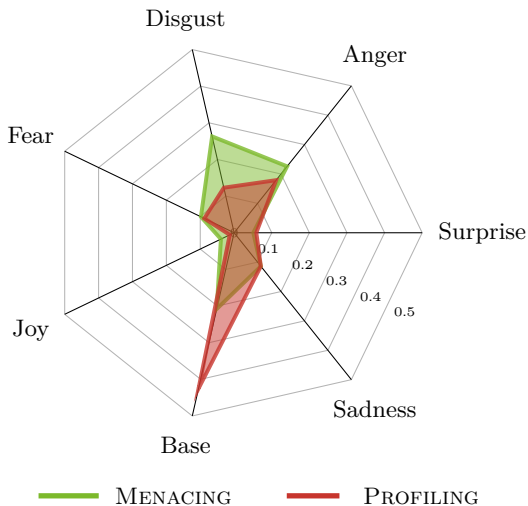
\begin{figure}[t!]
\centering
\begin{tikzpicture}
  \path (0,0) coordinate (O); 
  \foreach \X in {1,...,\D}{
    \draw (\X*\A:0) -- (\X*\A:\R);
  }

  \foreach \Y in {0,...,5}{
    \foreach \X in {1,...,\D}{
      \path (\X*\A:\Y*\R/5) coordinate (D\X-\Y);
    }
    \draw [opacity=0.3] (0:\Y*\R/5) \foreach \X in {1,...,\D}{
        -- (\X*\A:\Y*\R/5)
    } -- cycle;
    \pgfmathsetmacro{\label}{0.1*\Y}
    \node[font=\tiny] at ({335}:{\Y*\R/5 + 0.1}) {\tiny \pgfmathprintnumber[fixed, precision=1]{\label}};
`}

  \path (\A:\L) node (L1) {\normalsize Anger};
  \path (2*\A:2.9) node (L2) {\normalsize Disgust};
  \path (3*\A:\L) node (L3) {\normalsize Fear};
  \path (4*\A:\L) node (L4) {\normalsize Joy};
  \path (5*\A:2.9) node (L5) {\normalsize Base};
  \path (6*\A:\L) node (L6) {\normalsize Sadness};
  \path (7*\A:3.3) node (L7) {\normalsize Surprise};

  \fill[ncsuGreen, opacity=0.5]
    (\A:0.452*\R) 
    -- (\A*2:0.526*\R) 
    -- (\A*3:0.196*\R) 
    -- (\A*4:0.078*\R) 
    -- (\A*5:0.416*\R) 
    -- (\A*6:0.228*\R) 
    -- (\A*7:0.102*\R) 
    -- cycle;
  \draw [color=ncsuGreen, line width=1.5pt, opacity=0.8]
    (\A:0.452*\R) 
    -- (\A*2:0.526*\R) 
    -- (\A*3:0.196*\R) 
    -- (\A*4:0.078*\R) 
    -- (\A*5:0.416*\R) 
    -- (\A*6:0.228*\R) 
    -- (\A*7:0.102*\R) 
    -- cycle;

  \fill[ncsuRed, opacity=0.5]
    (\A:0.358*\R) 
    -- (\A*2:0.246*\R) 
    -- (\A*3:0.176*\R) 
    -- (\A*4:0.026*\R) 
    -- (\A*5:0.850*\R) 
    -- (\A*6:0.228*\R) 
    -- (\A*7:0.118*\R) 
    -- cycle;
  \draw [color=ncsuRed, line width=1.5pt, opacity=0.8]
    (\A:0.358*\R) 
    -- (\A*2:0.246*\R) 
    -- (\A*3:0.176*\R) 
    -- (\A*4:0.026*\R) 
    -- (\A*5:0.850*\R) 
    -- (\A*6:0.228*\R) 
    -- (\A*7:0.118*\R) 
    -- cycle;

  \begin{scope}[shift={(-2.8,-3.5cm)}]
    \draw[color=ncsuGreen, line width=2pt] (0.2,0) -- (1,0);
    \node[anchor=west] at (1.1,0) {{\normalsize \textsc{Menacing}}};
    
    \draw[color=ncsuRed, line width=2pt] (3.5,0) -- (4.2,0);
    \node[anchor=west] at (4.4,0) {{\normalsize \textsc{Profiling}}};
  \end{scope}
\end{tikzpicture}
\caption{Emotion reported in {\Menacing} and {\Profiling} reviews, displayed by a proportion of how frequent each emotion is in each category for \np{115581} reviews.}
\label{fig:emotion}
\end{figure}

We applied Hartmann's Emotion English DistilRoBERTa-base emotion classifier\footnote{Emotion English DistilRoBERTa-base: https://huggingface.co/j-hartmann/emotion-english-distilroberta-base/} to obtain the seven emotions of reviews indicating harassment---see Figure~\ref{fig:emotion}. The emotion classifier uses \emph{Neutral} which we rename to \emph{Base}. Figure~\ref{fig:emotion} displays the occurrence of these emotions in {\Menacing} and {\Profiling} reviews. We randomly selected and checked 300 reviews for the accuracy of each of the emotions.  

The disgust emotion is twice as prevalent in {\Menacing} as in {\Profiling} reviews. This difference can be attributed to {\Menacing} including unsolicited messages, intimate images, and bullying, which provoke greater disgust than stalking and similar offenses that constitute {\Profiling}.

The base emotion is twice as prevalent in {\Profiling} as in {\Menacing}. The base emotion correlates with users reporting fake profiles. Such reviews are often brief and seem to involve reduced engagement.

The surprise emotion is typically negative. Conversely, joy correlates with reviews highlighting the positive aspects of the app before mentioning any harassment. For instance, the review ``I love it but really laggy and too sexual'' shows how the user praises the app while acknowledging harassment. This observation suggests that positive emotion reviews do not preclude the presence of negative experiences. Therefore, it is crucial to consider the nuanced nature of user feedback when mining reviews and model training.

Findings~\ref{finding:RQ2a} and~\ref{finding:RQ2b} are the main outcomes for RQ$_2$ on the distribution of emotions.

\begin{mpsFinding}[finding:RQ2a]{Prevalence of Disgust}
Disgust is twice as frequent in {\Menacing} than in {\Profiling} reviews, reflecting strong negative reactions to unsolicited intimate content and bullying. {\Profiling} reviews contain double the proportion of neutral (base) emotions, often linked to fake profiles.
\end{mpsFinding}

Yi et al. \citep{yi2025detectingharassmentdefamationcyberbullying} map harassment to anger and disgust. By distinguishing {\Menacing} from {\Profiling}, we show how these forms of harassment have distinct emotional impacts.

\subsection{Negative vs. Neutral Reviews}
We considered the polarity of reviews based on their ratings. Reviews given 3-star ratings were considered neutral and reviews given 1 or 2-star ratings were considered negative. 

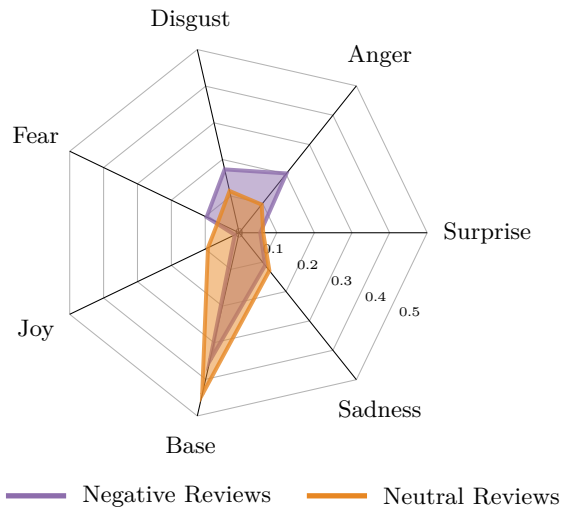
\begin{figure}[t!]
\centering
\begin{tikzpicture}
  \path (0:0cm) coordinate (O); 
  \foreach \X in {1,...,\D}{
    \draw (\X*\A:0) -- (\X*\A:\R);
  }

  \foreach \Y in {0,...,5}{
    \foreach \X in {1,...,\D}{
      \path (\X*\A:\Y*\R/5) coordinate (D\X-\Y);
    }
    \draw [opacity=0.3] (0:\Y*\R/5) \foreach \X in {1,...,\D}{
        -- (\X*\A:\Y*\R/5)
    } -- cycle;
    \pgfmathsetmacro{\label}{0.1*\Y}
    \node[font=\tiny] at ({335}:{\Y*\R/5 + 0.1}) {\tiny \pgfmathprintnumber[fixed, precision=1]{\label}};
`}

  \path (\A:\L) node (L1) {\normalsize Anger};
  \path (2*\A:2.9) node (L2) {\normalsize Disgust};
  \path (3*\A:\L) node (L3) {\normalsize Fear};
  \path (4*\A:\L) node (L4) {\normalsize Joy};
  \path (5*\A:2.9) node (L5) {\normalsize Base};
  \path (6*\A:\L) node (L6) {\normalsize Sadness};
  \path (7*\A:3.3) node (L7) {\normalsize Surprise};

  \fill[ncsuPurple, opacity=0.5]
    (\A:0.404*\R) -- (\A*2:0.346*\R) -- (\A*3:0.192*\R) -- (\A*4:0.026*\R) -- (\A*5:0.698*\R) -- (\A*6:0.222*\R) -- (\A*7:0.112*\R) -- cycle;
  \draw [color=ncsuPurple,line width=1.5pt, opacity=0.8]
    (\A:0.404*\R) -- (\A*2:0.346*\R) -- (\A*3:0.192*\R) -- (\A*4:0.026*\R) -- (\A*5:0.698*\R) -- (\A*6:0.222*\R) -- (\A*7:0.112*\R) -- cycle;

  \fill[ncsuOrange, opacity=0.5] 
    (\A:0.192*\R) -- (\A*2:0.228*\R) -- (\A*3:0.126*\R) -- (\A*4:0.184*\R) -- (\A*5:0.884*\R) -- (\A*6:0.258*\R) -- (\A*7:0.128*\R) -- cycle;
  \draw [color=ncsuOrange,line width=1.5pt, opacity=0.8]
    (\A:0.192*\R) -- (\A*2:0.228*\R) -- (\A*3:0.126*\R) -- (\A*4:0.184*\R) -- (\A*5:0.884*\R) -- (\A*6:0.258*\R) -- (\A*7:0.128*\R) -- cycle;

  \begin{scope}[shift={(-3.1,-3.5cm)}]
    \draw[color=ncsuPurple, line width=2pt] (0,0) -- (0.8,0);
    \node[anchor=west] at (0.9,0) {{\normalsize Negative Reviews}};
    
    \draw[color=ncsuOrange, line width=2pt] (4,0) -- (4.8,0);
    \node[anchor=west] at (4.9,0) {{\normalsize Neutral Reviews}};
  \end{scope}
\end{tikzpicture}
\caption{Distribution of emotions across negative (1-star) and neutral (2 and 3-star) reviews. Negative reviews exhibit anger, disgust, and fear two times more than neutral reviews. However, joy is nine times more prominent in neutral reviews than negative ones.}
\label{fig:spider_negative_neutral}
\end{figure}

First, we focus on emotions as shown in Figure~\ref{fig:spider_negative_neutral}. The proportions of negative and neutral ratings for emotions of base, sadness, and surprise stay constant, the most proportion differences occur in anger, disgust, fear, and joy.

Negative joy reviews are often sarcastic with reviews like ``\ldots use it if you wanna get kidnapped so have some fun people'' and ``This app is filled with groomers :D\ldots''. Reviewers use a sarcastic tone to express their frustration with the app. 

\begin{mpsFinding}[finding:RQ2b]{Victim Emotions}
Negative reviews exhibit anger, disgust, and fear approximately twice as often as neutral reviews, highlighting the emotional impact on victims.
\end{mpsFinding}

Neutral reviews for all emotions usually praise the app and explain online harassment like ``It’s a fun app but probably not as safe as it is fun\ldots'' and ``\dots love the concept but \ldots [a]t least 35 to 40 percent of people are just rude''.

Over 90\% of joy reviews are rated neutral because users who experience joy rate the app more positively. The opposite occurs for users who experience anger, disgust, or fear. Such users rate the app negatively, resulting in over 50\% of these reviews being rated as negative.

For female and male abusers, the proportions of neutral and negative reviews are equal. In neutral reviews, gender is mentioned when a user is disturbed by abusers. However, the user does not blame the app as seen in this example---``amazing app, but too many female scammers, trynna sell their body\ldots''.

In negative reviews explicitly mentioning gender, the user is disappointed with the app and their encounters with abusers as seen in this example---``[app] is disappointing at best! \ldots All of the girls seem fake and all they do is ask for money! \ldots''.

\section{Apps with Notable Harassment Incidents}
We analyzed the top 100 negatively and neutrally reviewed social apps to find those with over 50 harassment reviews. We found that 49 out of 100 Apple apps and 71 out of 100 Google apps met this threshold. 

\begin{mpsFinding}[finding:notableharassment]{Prevalence of Harassment}
The widespread presence of harassment in social apps is indicated by 49 out of the top 100 Apple apps and 71 out of the top 100 Google apps having over 50 harassment reviews.
\end{mpsFinding}

Table~\ref{tab:top_harassment} shows apps with over 500 cases of harassment, some of which appear in both Apple and Google. Our computational model handles social app reviews. Among all social apps that we identify, 33 out of 37 apps from Google and all 11 apps from Apple are dating or chatting apps.

\begin{mpsFinding}[finding:]{Harassment Across App Types}
The top 10 apps with the highest harassment reported across both Google and Apple app stores are chatting and dating apps. 
\end{mpsFinding}

Example~\ref{box:popular_app_harassment_examples} displays reviews illustrating harassment found on MeetMe (chatting app), Once (dating app), and Plenty of Fish (dating app). Reviews M1, O1, and P1 focus on the extensive number of predators and M2, O2, and P2 on fake profiles.

\begin{mybox}[box:popular_app_harassment_examples]{Harassment on popular apps.}

\textbf{MeetMe}\\
\textbf{M1:} ``Stay away unless you want to interact with absolute scum. Meetme does nothing to stop the horde of potential child/sexual predators.'' \\ \\
\textbf{M2:} ``90\% are fake profiles\ldots Trying to scam or sell you content''\\ 

\textbf{Once} \\
\textbf{O1:} ``This app is full of weirdos creeps and fake accounts it also doesnt let me screen shot anything or save pictures that are sent to me'' \\ \\
\textbf{O2:} ``Absolutely ZERO REAL PROFILES Dont waste a second on this bs Again it is LITERALLY ALL FAKE PROFILES''\\ 

\textbf{Plenty of Fish} \\
\textbf{P1:} ``This app is full of prostitution If you get it do not give away any of your information'' \\ \\
\textbf{P2:} ``Pof stands for plenty of fakes Fakes and scammers is all thats on this site''
\end{mybox}

Let us focus on user expectations of chatting and dating apps. Example~\ref{box:social-app-culture} contains examples of what victims expected when they installed the dating or chatting app and what they were faced with instead. The app IDs correspond to app IDs in Table~\ref{tab:top_harassment_google} and Table~\ref{tab:top_harassment_apple}.

\begin{table*}[htbp]
    \caption{Social app harassment table displaying the app names along with critical harassment instances and the number of reviews indicating online harassment. Apps with over 500 instances of harassment are included.}
    \footnotesize
    \begin{subtable}{\textwidth}
    \centering
    \begin{tabular}{llrrr}
        \toprule
        \textbf{App Name} & \textbf{Harassment Types} & \textbf{Total} &  \textbf{\Menacing}  &  \textbf{\Profiling}  \\
        \midrule
        Once & stalking & \np{2876} & \np{201} & \np{2774} \\
        Plenty Of Fish & blackmail, pedophilia, stalking & \np{2232} & \np{456} & \np{2056} \\
        MeetMe & blackmail, pedophilia, stalking & \np{1984} & \np{614} & \np{1726} \\
        Zoosk & stalking & \np{1565} & \np{189} & \np{1436} \\
        Happn & blackmail, pedophilia & \np{1416} & \np{221} & \np{1331} \\
        Tinder & blackmail, pedophilia, stalking & \np{1332} & \np{231} & \np{1197} \\
        Skout & blackmail, pedophilia, stalking & \np{1314} & \np{431} & \np{1098} \\
        TanTan & blackmail & \np{1258} & \np{162} & \np{1202} \\
        Hily & blackmail, pedophilia, stalking & \np{1231} & \np{306} & \np{1029} \\
        Dating.com & pedophilia & \np{1214} & \np{110} & \np{1156} \\
        Hinge & blackmail, doxxing, pedophilia, stalking & \np{1159} & \np{268} & \np{955} \\
        OkCupid & blackmail, pedophilia, stalking & \np{1099} & \np{174} & \np{994} \\
        Badoo & blackmail, pedophilia, stalking & \np{1016} & \np{321} & \np{779} \\
        Bumble & blackmail, stalking & \np{912} & \np{228} & \np{745} \\
        Hoop & blackmail, pedophilia, stalking & \np{897} & \np{352} & \np{727} \\
        iFlirts & pedophilia & \np{830} & \np{78} & \np{794} \\
        Wink & blackmail, pedophilia, stalking & \np{791} & \np{253} & \np{659} \\
        MEEFF & blackmail, pedophilia, stalking & \np{785} & \np{336} & \np{584} \\
        Match Dating & stalking & \np{766} & \np{142} & \np{658} \\
        Tumblr & pedophilia, stalking & \np{764} & \np{641} & \np{189} \\
        Seeking & blackmail, pedophilia, stalking & \np{757} & \np{248} & \np{612} \\
        OfferUp & stalking & \np{723} & \np{118} & \np{643} \\
        RandoChat & blackmail, child abuse, pedophilia & \np{723} & \np{320} & \np{489} \\
        eHarmony & blackmail, pedophilia, stalking & \np{699} & \np{139} & \np{592} \\
        Nextdoor & doxxing, pedophilia, stalking & \np{673} & \np{530} & \np{174} \\
        Grindr & blackmail, pedophilia, stalking & \np{633} & \np{305} & \np{408} \\
        WooPlus & blackmail, stalking & \np{632} & \np{209} & \np{497} \\
        Taimi & blackmail, stalking & \np{627} & \np{367} & \np{301} \\
        Muzz & blackmail & \np{618} & \np{132} & \np{526} \\
        Sugo & blackmail & \np{609} & \np{170} & \np{511} \\
        OYO & blackmail & \np{573} & \np{90} & \np{501} \\
        Omegle & pedophilia, stalking & \np{556} & \np{282} & \np{314} \\
        CSL & blackmail, pedophilia & \np{554} & \np{106} & \np{520} \\
        Monkey & blackmail, pedophilia & \np{548} & \np{462} & \np{109} \\
        JAUMO & blackmail, pedophilia, stalking & \np{530} & \np{89} & \np{496} \\
        Woo & blackmail, stalking & \np{507} & \np{75} & \np{486} \\
        Waplog & blackmail & \np{506} & \np{91} & \np{465} \\
        \bottomrule
        \end{tabular}
        \vspace{0.5em}
        \subcaption[]{There are 37 apps from Google Play Store with critical harassment cases and over 500 harassment reviews.}
        \label{tab:top_harassment_google}
        \end{subtable}
        \\
        \begin{subtable}{\textwidth}
        \centering
        \begin{tabular}{llrrr}
        \toprule
        \textbf{App Name} & \textbf{Harassment Types} & \textbf{Total} &  \textbf{\Menacing}  &  \textbf{\Profiling} \\ \midrule
        MeetMe & blackmail, pedophilia, stalking & \np{1684} & \np{632} & \np{1406} \\
        Kinkoo & blackmail & \np{1384} & \np{318} & \np{1296} \\
        Hoop & blackmail, pedophilia, stalking & \np{1091} & \np{608} & \np{764} \\
        Wink & blackmail, pedophilia & \np{968} & \np{511} & \np{639} \\
        Yubo & blackmail, doxxing, pedophilia, stalking & \np{727} & \np{577} & \np{198} \\
        Once & blackmail & \np{680} & \np{50} & \np{652} \\
        Wizz & blackmail, pedophilia, stalking & \np{656} & \np{497} & \np{213} \\
        Addchat & blackmail, pedophilia & \np{655} & \np{430} & \np{310} \\
        Bustr & blackmail & \np{650} & \np{234} & \np{599} \\
        Whisper & blackmail, pedophilia, stalking & \np{616} & \np{459} & \np{339} \\
        Skout & blackmail, pedophilia, stalking & \np{530} & \np{205} & \np{424} \\
        \bottomrule
        \end{tabular}
        \vspace{0.5em}
        \subcaption[]{There are 11 apps from Apple App Store with critical harassment cases and over 500 harassment reviews.}
        \label{tab:top_harassment_apple}
        \end{subtable}
        
    \label{tab:top_harassment}
\end{table*}

\begin{mybox}[box:social-app-culture]{User expectations in reviews.} \
\textbf{Chatting apps}\\
``I downloaded this app to meet new people and talk to new people 90 percent of everyone on there are girls trying to sell there nudes and stuff smh this world is going in the wrong direction.'' (Wink@Apple)\\

``This is not at all what I expected. All the messages I got were from some guy/girl asking for nudes\ldots This app is full of nasty people\ldots'' (Yubo@Apple)\\

``So many fake profiles to be exact 451 spam messaging me just within the last 7 days.'' (MeetMe1@Google)\\

``For gods sake within half an hour of signing up I was inundated with men who clearly didnt read my profile that stated I was happily married and looking for like minded friends only and to not give me sexual offers. \ldots this is all sexual harassment\ldots'' (MeetMe2@Google)\\

\textbf{Dating apps}\\
``Terrible experience Mostly s3x workers and only fans girls trying to market their nudes.'' (Hily@Google)\\

``This app is all bots I was on it for 2 days and dont think I saw a real person.'' (Plenty of Fish@Google)\\

``I came on this app looking for a relationship. I met a girl and was chatting, but it was actually a man that tricked me into sending nudes and threatening me to send money or he will post it. Be careful of scammers on the app.'' (Kinkoo@Apple)

\end{mybox}

The patterns between chatting and dating apps in Example~\ref{box:social-app-culture} are similar. First, both chatting and dating apps are concerned about the abundance of fake profiles or bots on the app [Example~\ref{box:social-app-culture}: MeetMe2@Google), Plenty of Fish@Google, Kinkoo@Apple]. Users are primarily concerned about personal information privacy and monetary risk when downloading apps \citep{Jorgensen_Risk_in_Applications}. On social apps, particularly dating and chatting apps, users expect to interact with other humans but may encounter bots or ads.

The second similarity is abusers sending or selling intimate images and messages [Example~\ref{box:social-app-culture}: Wink@Apple, Yubo@Apple, MeetMe2@Google, Hily@Google]. 
When abusers send or market intimate images or messages, they do not have the victim's consent. 
Reviews Wink@Apple and Hily@Google detail how the users are disappointed by their experience since their expectations of conversing result in the other person marketing their intimate images. 
Phan et al. \citep{phan2021threaten} emphasize users' expectations of forming relationships through dating apps. However, reviews of these apps report rape, stalking, and child-exploitation, and other crimes.

Reviews of Wink@Apple and Hily@Google shed light on the culture of chatting and dating apps from the abuser's and victim's points of view. Abusers use these apps as platforms to promote their sex business, an opportunity not available on other apps. On the other hand, victims argue that the dissemination of intimate images should be prohibited on social platforms.

Although social apps share technical features such as the ability to post content and send messages, certain apps yield greater online harassment than others. What differentiates these apps are robust mechanisms to prevent misuse. 

\section{Advice for Developers} \label{sec:harassment_report} 

We identified several critical online actions of apps containing such instances in Table~\ref{tab:top_harassment}. Table~\ref{tab:top_harassment} includes apps where instances of harassment exceed 500 reported cases, highlighting patterns of misuse. Each entry includes the app's name, app ID, a list of critical harassment incidents (pedophilia, stalking, blackmailing, and doxxing), and the respective counts for {\Menacing}, {\Profiling}, and total reviews. 

A prominent observation from Table~\ref{tab:top_harassment} is the prevalence of blackmail, pedophilia, and stalking. Additionally, we identified cases of child abuse and doxxing.

Abusers can blackmail a victim by threatening to share the victim's intimate images or doxx them. Child abuse instances involve the abuser attempting to groom children by manipulating them to send personal pictures.

Our approach identifies the specific forms of abuse users experience and can help developers improve their mitigation. We contacted the developers of all 48 apps we identified, providing details on the harassment we uncovered. 
Three responded, and one of them was unwilling to address feedback. Jaumo\footnote{Jaumo: https://live.jaumo.com/en} (dating app) expressed openness to making changes but indicated resource constraints as a factor. Hinge\footnote{Hinge: https://hinge.co} (dating app) says they combat such malice but acknowledge their current methods may be inadequate.

Thus, developers are aware of harassment through their apps and recognize the need for stronger measures, but mitigation is limited by implementation challenges and resource constraints. We speculate that many developers, despite acknowledging these risks, do not prioritize addressing them sufficiently.

App stores can actively monitor reviews to stay ahead of emerging threats. Regular audits focused on online misbehavior may help ensure apps meet safety standards, with clear terms of service and enforcement for violations. Developers can implement stronger verification systems, such as identity verification to ensure that users are who they claim to be and content moderation to reduce harassment. User-friendly reporting features may also be in place to allow quick flagging and resolution of misconduct \citep{youth_trust}.

Users can continue to report harassment in app reviews to help highlight problems and push developers to act. By working together, platform stores and developers can prevent online harassment and create a safer digital environment as social apps continue to evolve.

\section{Conclusion}
\label{sec:conclusion}
Social apps are designed to facilitate human interaction. However, they can enable harmful behaviors if used irresponsibly. 
As developers create apps that support more sophisticated forms of social interaction, a sociotechnical systems perspective draws attention to how users (mis)behave through the app and thus uncovers threats that are not captured in the technology alone.

This study relies on app reviews, providing insights into how apps are misused for online harassment through users' firsthand experiences. 
Knowing about harassment and its various forms on their apps can help developers make course corrections in their offerings.  
Through such interventions, developers can prevent their apps from becoming toxic, thereby contributing to a more secure and user-friendly experience.

\subsection{Threats to Validity}

While extracting reviews from apps, we mitigated the threat of sampling bias by extracting recommended apps as described in Section~\ref{sec:dataset}. 
During model training, we introduced reviews without keywords to mitigate keyword bias, as explained in Section~\ref{sec:dataset}. Since developers continually update apps, we include only reviews posted January 1, 2020 onward. 
However, some threats remain. Some reviews could be fake. 
Further, only the problems identified by the reviewers are included. 

We excluded reviews in languages other than English, so our findings may not generalize across languages and cultures. 
We did not analyze positive reviews, which may have limited our ability to compare the emotion analysis and reporting patterns across all user ratings.

Our model may miss new apps, whose reviews don't yet indicate harassment. A possible resolution would be to weight the classification towards newer reviews. 
Also, sometimes harassment may be triggered through nonsocial apps, such as Google Maps; such cases are out of our scope. 

Our analysis of the abusers' gender is based on the \np{18459} reviews that mention gender, and may be biased relative to the entire dataset.

\subsection{Future Work}

Future research could expand on our findings in two ways. A multilingual analysis could provide deeper insights into harassment patterns across different cultures, highlighting variations in online abuse. Additionally, it could offer a more comprehensive understanding of user feedback for developers, helping to identify cultural differences to improve app functionality.

Our work focuses on app store reviews. Future work could explore additional aspects such as in-app functionalities, app permissions, privacy policies, and user interviews to further validate and expand our findings. Additionally, the scope of this work can be expanded to include non-social apps to better understand online harassment.

\subsection{Ethical Considerations}

Some ethical concerns arise from this work. Since we analyzed public reviews of publicly available apps, author consent was not required. To preserve anonymity, we removed usernames and any personally identifiable information from both our dataset and this manuscript. App reviews often contain personal accounts of online harassment, sometimes written by victims, which could be distressing for individuals with similar experiences. To mitigate potential harm, we labeled the reviews ourselves instead of using crowd workers.

\section*{Acknowledgments}
Thanks to the Department of Defense for partially supporting this research under the Science of Security Lablet at NC State University.

\begin{IEEEbiography}{Sanjana Cheerla}{\,} is a PhD Student in Computer Science at NC State University. Her interests include NLP, biomedical relation extraction, and LLMs. Contact her at scheerl@ncsu.edu.
\end{IEEEbiography}

\begin{IEEEbiography}{Vaibhav Garg}{\,} is the Collegiate Assistant Professor at Virginia Tech. His research interests include NLP, computational social science, and online privacy. Contact him at vaibhavg@vt.edu.
\end{IEEEbiography}

\begin{IEEEbiography}{Saikath Bhattacharya}{\,} is an Assistant Professor of Cybersecurity at the School of Information Technology, Illinois State University. His interests include software reliability, cybersecurity, and systems engineering. He received his Ph.D. in computer engineering from the University of Massachusetts Dartmouth.   Contact him at sbhatt8@ilstu.edu.
\end{IEEEbiography}

\begin{IEEEbiography}{Munindar P.~Singh}{\,} is the SAS Institute Distinguished Professor of Computer Science at NC State University. His interests include the engineering and governance of sociotechnical systems, and AI ethics. Singh is a Fellow of AAAI, AAAS, ACM, and IEEE. Contact him at singh@ncsu.edu.
\end{IEEEbiography}

\end{document}